\documentclass{pasj00}
\begin{document}
\SetRunningHead{Y. Uchiyama, T. Takahashi, and F. A. Aharonian}{X-ray Emission from the Molecular Cloud Associated with SNR RX~J1713.7$-$3946}
\Received{2002/02/28}%{yyyy/mm/dd}
\Accepted{2002/08/02}%{yyyy/mm/dd}

\title{Flat Spectrum X-ray Emission from the Direction 
of a Molecular Cloud Associated with SNR RX~J1713.7$-$3946}

%%% begin:list of authors
\author{Yasunobu \textsc{Uchiyama}
      and Tadayuki \textsc{Takahashi}}
\affil{The Institute of Space and Astronautical Science,
3-1-1 Yoshinodai, Sagamihara, Kanagawa 229-8510 \\
and
Department of Physics, University of Tokyo, 
7-3-1 Hongou, Bunkyo, Tokyo 113-0033}
\email{uchiyama@astro.isas.ac.jp}
\and
\author{Felix A. \textsc{Aharonian}}
\affil{Max-Planck-Institut f\"{u}r Kernphysik, Postfach 103980,
D-69029 Heidelberg, Germany}

%%% end:list of authors

%% `\KeyWords{}' always has to be placed before `\maketitle'.
 %%%Do NOT move this preamble from here!

\maketitle

\begin{abstract}
We report on the discovery of a diffuse X-ray source with ASCA,
presumably associated with a molecular cloud in the vicinity of 
the supernova remnant RX~J1713.7$-$3946.
The energy spectrum (1--10 keV) of the hard X-ray source shows 
a flat continuum, which is described by a power-law with a photon 
index of $\Gamma = 1.0^{+0.4}_{-0.3}$. We argue that this unusually 
flat spectrum can be best interpreted  in terms of characteristic  
bremsstrahlung emission from the loss-flattened distribution of 
either sub-relativistic 
protons or mildly relativistic electrons. The strong shock of 
RX~J1713.7$-$3946, which is likely to interact with 
the molecular cloud, as evidenced by CO-line observations,  
seems to be a natural site of acceleration of such nonthermal particles.
The observed luminosity of 
$L_{\rm X} = 1.7 \times 10^{35} $ erg~s$^{-1}$ (for a distance of 6 kpc) 
seems to require a huge kinetic energy
of about $10^{50}$ erg in the form of nonthermal 
particles to illuminate the cloud. 
The shock-acceleration at RX~J1713.7$-$3946 can barely 
satisfy this energetic requirement, unless 
(i) the source is located much closer than the preferred distance of 6 kpc 
and/or (ii) the mechanical energy of the supernova
explosion essentially exceeds $10^{51}$ erg.
Another possibility would be that an essential 
part of the lost energy due to the ionization and heating of gas, 
is somehow converted to plasma waves, which return this energy 
to nonthermal particles through their turbulent reacceleration 
on the plasma waves. 
\end{abstract}

\section{Introduction}

Supernova remnants (SNRs) are commonly believed to be 
major sites for the production of Galactic cosmic rays.
The diffusive shock acceleration mechanism 
naturally accounts for the hard power-law production 
spectra of cosmic rays in their sources, but, as long as the 
particle injection rate remains an unsolved problem, 
the theory does not tell us conclusively what fraction of 
the initial kinetic energy of an explosion can be 
transferred to cosmic rays (see e.g. the recent review by \cite{Malkov01}).
Therefore, it is extremely important to  derive    
the total energy contained in  subrelativistic and 
relativistic  particles from  observations of the relevant components
of nonthermal electromagnetic radiation. 

Radio synchrotron emission provides information about  GeV electrons 
in most shell-type SNRs.  Synchrotron X-ray emission recently  discovered 
in SN~1006 \citep{Koyama95}  and some other shell-type SNRs  indicates  
that the electron acceleration continues effectively up to multi-TeV energies.
It is believed, however, that the accelerated protons constitute the major 
fraction of nonthermal energy to which the mechanical energy of a supernova 
explosion is transferred  through shock acceleration. The best way to 
explore the proton component of accelerated particles in SNRs is the detection 
of  gamma-rays by the decay of $\pi^0$ mesons produced in collisions between 
cosmic-ray protons and ambient matter \citep{Drury94, Naito94}. 
The detection of gamma-rays of hadronic origin from SNRs  by 
current and planned space- and ground-based instruments requires 
high-density environments  in or close to the particle-acceleration sites.  
Large molecular clouds interacting with shells of SNRs may act as
effective gas targets for the production of $\pi^0$ mesons and 
their subsequent decay to high-energy $\gamma$-rays \citep{ADV94}. 
It has recently been claimed \citep{Butt01} 
that there exists a compelling positional association of 
the unidentified $\gamma$-ray 
source  3EG J1714$-$3857 with a dense molecular cloud (cloud A)  
overtaken, most probably,
by  the shock front  of  the SNR RX~J1713.7$-$3946 (G 347.3$-$0.5).
Moreover, it  has been argued that the TeV gamma-rays from this SNR
detected by the CANGAROO collaboration are also the result of  interactions 
of accelerated protons with nearby dense gas targets \citep{Enomoto02},
though \citet{Reimer02} claimed that 
this interpretation seems to be inconsistent with EGRET observations of this
region.

The flux of the subrelativistic component of the accelerating protons
is generally inconclusive both on observational and theoretical grounds, 
but it is quite possible that this component dominates the total energetics
of nonthermal particles. Unfortunately, subrelativistic protons arriving 
at the Earth are not a representative sample of the cosmic-ray flux  in the 
Milky Way owing to the effect of solar modulation. 
Ionization losses of 
these particles  play an important role in heating both diffuse neutral gas 
and molecular clouds, and in generating free electrons in molecular clouds, 
which  are crucial for interstellar chemistry.
It is therefore of great importance  to measure the energy released in the 
form of  subrelativistic particles  by cosmic-ray accelerators.

The ideal diagnostic tool to probe subrelativistic cosmic rays 
is the nuclear prompt $\gamma$-ray line emission  
at MeV energies \citep{Ramaty79}.
However,  conservative estimates show that the fluxes of 
even  most prominent $\gamma$-ray lines  are only marginally detectable 
because of limited sensitivities of   gamma-ray instruments in the MeV band.  
On the other hand, as we discuss below, 
the search for the sites of concentrated subrelativistic protons 
using bremsstrahlung X-rays could be very promising
with the help of superior sensitivities of current X-ray detectors,
even though only a small fraction ($\sim 10^{-5}$) of the nonthermal 
energy of subrelativistic protons and electrons is radiated away 
in the form of bremsstrahlung X-rays.  
Remarkably, if the subrelativistic cosmic rays 
reside  in a dense gas environment, the ionization losses would result in 
a significant flattening of the low-energy spectra, and thus leading to 
the  characteristic ``$1/\varepsilon$''  type bremsstrahlung X-ray spectrum 
\citep{Uchiyama02}.  Such an unusually hard X-ray spectrum may serve as 
a distinct signature of  nonthermal bremsstrahlung origin of X-ray emission.

In our previous paper \citep{Uchiyama02} we reported on
the detection of hard X-rays from a localized region 
in the SNR $\gamma$~Cygni,  which we attributed to $1/\varepsilon$ 
bremsstrahlung  from the loss-flattened electron distribution. 
In this paper we present a more pronounced  example of such hard X-radiation
arriving  from a massive cloud in the vicinity of  SNR RX~J1713.7$-$3946.

\section{SNR RX~J1713.7$-$3946} \label{sec:rxj}

\subsection{Basic Information from Previous Studies} 

The shell-type SNR RX~J1713.7$-$3946 (G 347.3$-$0.5),
which has an angular radius of about $40\arcmin$,
was first discovered during the ROSAT 
All-Sky Survey \citep{Pfeffermann96}.  The detection  of  synchrotron
X-ray emission  with ASCA 
\citep{Koyama97, Slane99}  revealed  that
the remnant's shell is a presently operating
source of multi-TeV electrons, which are responsible for 
the synchrotron X-rays.  Moreover,   
the TeV $\gamma$-rays  detected by the CANGAROO collaboration from  
this supernova remnant \citep{Muraishi00,Enomoto02} provide direct evidence 
of the acceleration  of particles -- electrons and/or protons -- to energies 
well beyond 10 TeV.  
This SNR is very faint at radio wavelengths.
Based on a lack of thermal X-ray emission, the supernova shock 
is considered to be expanding in a low-density environment, 
such as a stellar-wind cavity \citep{Slane99}. 

The distance $d$, age $\tau$, and the 
explosion energy  $E_{51}=E/10^{51}$ erg are 
three important parameters for understanding
the origin of nonthermal radiation components.  
At the perimeter of RX~J1713.7$-$3946,
there is  a dense  ($n \sim 1 \times 10^{3}$~cm$^{-3}$) molecular cloud 
(cloud~A) that seems to have 
a very high ratio of two rotational transition lines of CO molecules,
CO($J$=2--1)/CO($J$=1--0). This  
suggests an interaction between cloud~A  and
the SNR RX~J1713.7$-$3946 \citep{Slane99, Butt01}.
The distance to cloud A ($V_{\rm LSR}$= $-$94 km s$^{-1}$)
was kinematically estimated to be $6.3\pm 0.4$ kpc by using the 
standard Galactic rotation curve \citep{Slane99}. Thus, if the 
molecular cloud and the supernova remnant 
are  indeed \emph{physically} associated,
the distance to RX~J1713.7$-$3946 could be significantly 
larger than the original estimate of about 1 kpc based on 
an absorbing column-density value \citep{Koyama97}. 
For the preferred distance to the source
of about 6 kpc, 
a reasonable set of the age and the explosion energy of RX~J1713.7$-$3946 
is in the ranges 18700--40600 yr and $E_{51}=$2.2--1.7  
\citep{Slane99}. 
However, a different solution of a distance $d=2 \ \rm kpc$,
with the remnant's age of 2100--3400 yr and 
the SN explosion energy of
$E_{51}=$1.4--2.2, cannot be ruled out.
The smaller distance and correspondingly 
younger remnant agree better with the  hypothesis  of    
\citet{Wang97}  who argued that RX~J1713.7$-$3946 
could be a likely counterpart 
of the guest star of 393~\textsc{a.d.} in the constellation Scorpius.

\subsection{ASCA Analysis} \label{sec:asca}

\begin{figure}[t]
  \begin{center}
    \FigureFile(80mm,80mm){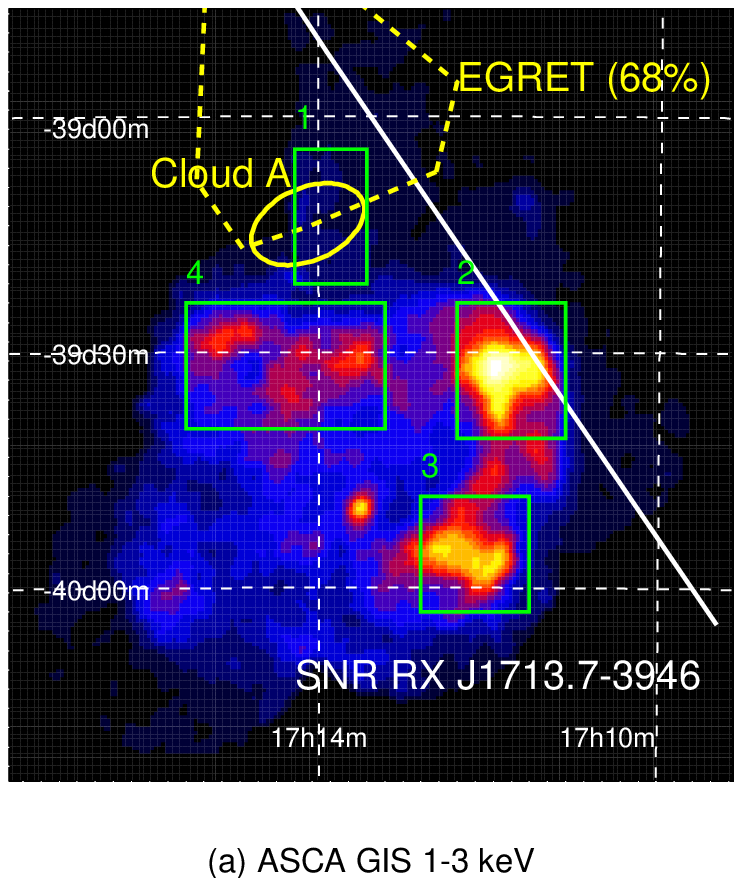}
    \FigureFile(80mm,80mm){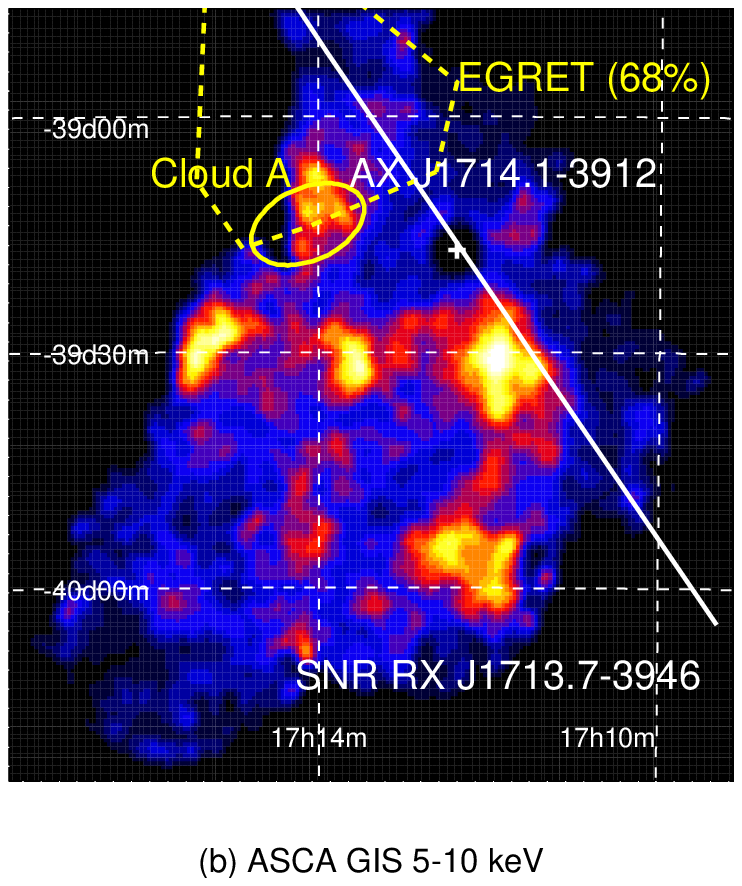}
    %%% \FigureFile(width,height){filename}
  \end{center}
  \caption{X-ray surface brightness (color map linearly spaced from 
black to white) of the SNR RX~J1713.7$-$3946 
in (a) 1--3 keV and (b) 5--10 keV energy bands
obtained with the ASCA GIS detectors.
Right ascension and declination are in the J2000 coordinate system.
The white thick line represents the Galactic plane.
A transient X-ray source \citep{Koyama97} 
is blanked out [a white cross in (b)].
Depicted are the 68\% confidence error contour of the EGRET 
unidentified source 3EG J1714$-$3857 \citep{Butt01}, and
an approximate outline of cloud~A 
based on the CO($J$=1--0) intensity \citep{Bronfman89}.}
\label{fig:ximage}
\end{figure}

The SNR RX~J1713.7$-$3946 was partially observed with 
the ASCA satellite on 1996 September,
and most of the remnant was covered on 1997 March.
The data were screened by standard procedures, 
except that strict rise-time screening was not applied 
to the data taken on 1997 March.
Below we present the results from the Gas Imaging Spectrometer (GIS) 
by adding two identical detectors (GIS~2 and GIS~3).

Figure \ref{fig:ximage}  shows the X-ray images of the SNR RX~J1713.7$-$3946
in the soft 1--3 keV and hard 5--10 keV energy bands.
After subtracting the instrumental background and 
correcting vignetting effect, images are smoothed 
with a Gaussian of standard deviation of $\timeform{0.75'}$ 
and $\timeform{1.0'}$ for the soft and hard band, respectively.
As it has been noted by previous studies \citep{Koyama97,Slane99},
shell-like synchrotron X-ray images  (region 2--4) are seen 
in both energy bands. In addition to region 2--4, the hard-band image 
clearly shows the existence of an extended source (region 1) 
being coincident with cloud~A, although it is barely seen in the soft band 
because of diffuse X-ray emission along the Galactic plane.
Approximate center of this new source is 
at a celestial coordinate (J2000) of 
R.A.=$\timeform{17h14m7.0s}$ and 
Decl.=$\timeform{-39D12'12.0''}$.
We thus refer to this source as AX~J1714.1$-$3912.
The count rate in the 5--10 keV band of AX~J1714.1$-$3912 is 0.029 c~s$^{-1}$,
which corresponds to a 34~$\sigma$ statistical significance in excess of 
the background in this region. 
The total number of source photons detected in the 1--10 keV band is 
derived to be 3340.
Unfortunately, the source lies very near to the edge of
the observed field. It is therefore difficult to determine 
any extent of the source, which is distorted by the off-axis broad 
point spread function of the telescope.

\begin{figure}
  \begin{center}
  \FigureFile(80mm,50mm){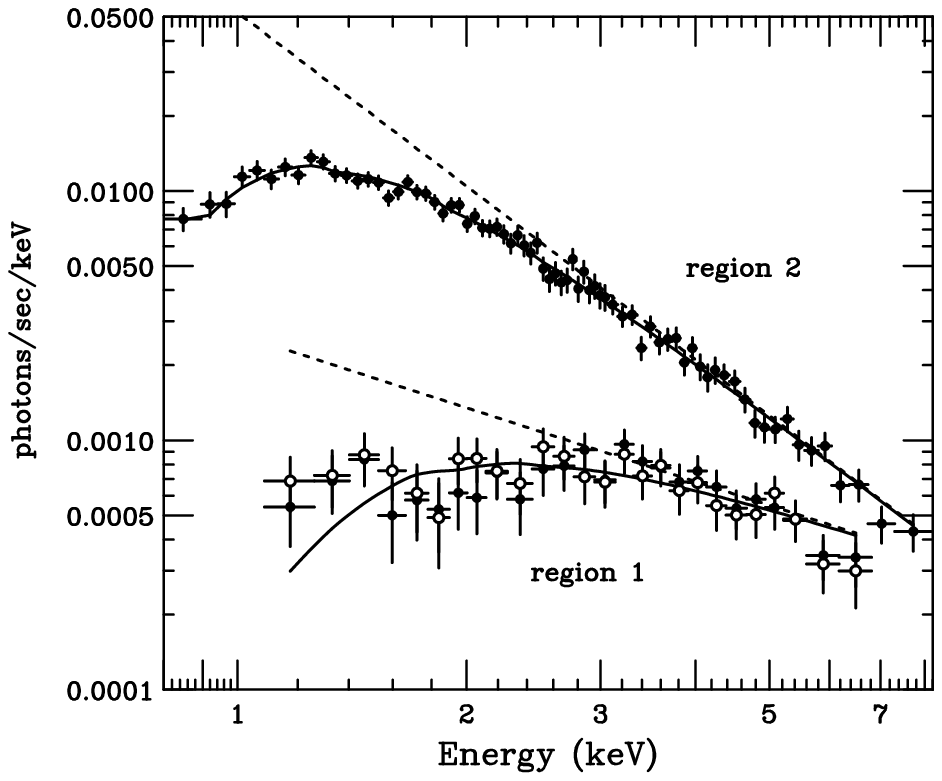}
    %%% \FigureFile(width,height){filename}
  \end{center}
  \caption{
Unfolded X-ray spectra of region~1 (AX~J1714.1$-$3912) and region~2 
measured by the ASCA GIS detectors.
The solid lines are the best-fit power-law models, taking account of
the photoelectric absorption at lower energies 
(see table \ref{tbl:fit}),
while the dashed lines are absorption-corrected power-law spectra.
For illustrative purposes,
we show the X-ray spectra of region 1 being derived  by
using an alternative background field (open circles).}
\label{fig:spectra}
\end{figure}

The X-ray spectrum of each region (1--4) was derived 
by subtracting the contributions from the Galactic diffuse emission
taken from a nearby field, and
then fitted with a photoelectric-absorbed power-law model,
described as
$I(\varepsilon) \propto \varepsilon^{-\Gamma} 
\exp \left[ - N_{\rm H} \sigma(\varepsilon) \right]$,
where $\Gamma$ is a photon index,
$\sigma(\varepsilon)$ is the photoelectric cross-section, 
and $N_{\rm H}$ is the equivalent hydrogen column density.
All four spectra were well-fitted by these simple models. 
The best-fit parameters and their 90\% confidence limits 
are summarized in table~\ref{tbl:fit}. 
The energy spectra and the best-fit models
of region~1 (i.e., AX~J1714.1$-$3912) and region~2 are shown 
in figure~\ref{fig:spectra}.
The power-law indices in region 2--4, $\Gamma =$ 2.1--2.3, 
are consistent with previous measurements 
and are best explained by synchrotron radiation
from ultrarelativistic electrons.
We found that the spectrum of AX~J1714.1$-$3912 is best fitted by 
an extremely flat power-law model with $\Gamma = 1.0^{+0.4}_{-0.3}$.
Within the statistical uncertainties,
the absorbing column density for region 1 agrees with 
that obtained for region 2--4. 
Attempts at fitting a thermal-bremsstrahlung model 
to the region~1 spectrum 
gave a lower-limit temperature of 35 keV (90\% confidence),
which is unreasonably high for SNRs.

\begin{table}
  \caption{Spectral fits to the ASCA data.}\label{tbl:fit}
 \footnotesize
  \begin{center}
    \begin{tabular}{lcccc}
        \hline \hline
Region & $N_{\rm H}$ & $\Gamma$ & $F_{1-10\ \rm keV}$ 
 & $\chi^2_{\nu}(\nu)$ \\
 & $(10^{22}\rm cm^{-2})$ & & $(10^{-11}\rm erg\ cm^{-2}s^{-1})$ 
 &  \\
        \hline
region 1  & $1.28^{+1.00}_{-0.70}$ & $0.98^{+0.44}_{-0.34}$ 
& $4.0^{+0.4}_{-0.3}$ & 1.11(23) \\
\\[0mm]
region 2  & $0.68^{+0.07}_{-0.06}$ & $2.31\pm{0.07}$ 
& $13.7\pm{0.4}$ & 0.72(71) \\
region 3  & $0.69\pm{0.05}$ & $2.29\pm{0.06}$ 
& $11.0\pm{0.3}$ & 0.98(87) \\
region 4  & $0.49\pm{0.05}$ & $2.13^{+0.06}_{-0.05}$ 
& $15.5\pm{0.3}$ & 0.72(85) \\
        \hline
    \end{tabular}
  \end{center}
Note. --- Best-fit parameter values and 90\% confidence intervals;
$N_{\rm H}$ is equivalent hydrogen column density,
$\Gamma$ is photon index, and
$F_{1-10\ \rm keV}$ is the unabsorbed flux in the 1--10 keV band.
$\chi^2_{\nu}$ is the reduced chi-square for the best-fit model and
$\nu$ is the number of degrees of freedom of the fit.
Region 1--4 are schematically indicated by the boxes 
in figure~\ref{fig:ximage}.
\end{table}

\section{Discussion} \label{sec:discuss}

\subsection{On the Origin of Flat X-Ray Emission}

We have discovered a new X-ray source, AX~J1714.1$-$3912,
with an extent of $\sim 10\arcmin$,
characterized by an extremely 
flat power-law with $\Gamma = 1.0^{+0.4}_{-0.3}$.
The luminosity of the source is estimated to be
$L_X = 1.7 \times 10^{35} d_6^2$ erg~s$^{-1}$,
where $d_6$ is the distance  to the source normalized to 6 kpc.
Since the luminosity increases linearly with a high-energy cutoff
for the $\varepsilon^{-1}$ type spectrum, 
the above estimate should be considered as a lower limit 
for the luminosity.
It cannot be the radiation of thermal gas, 
unless we assume an extremely hot optically 
thin plasma with a temperature significantly exceeding 10 keV.  
Since the formation of such a thermal source seems to be quite 
problematic,  given  its extended character,  
below we assume that the X-radiation has a nonthermal origin.  
AX~J1714.1$-$3912 has a good spatial association with cloud A 
(see figure~\ref{fig:ximage}). 
In the following discussion, therefore, we assume that 
the X-ray emission actually comes from cloud A. If so, 
this would be a new striking nonthermal phenomenon in molecular clouds.

Nonthermal particles that are illuminating cloud A in the X-ray band 
could be supplied by either internal or external accelerators.
A potential  candidate for an external accelerator is the strong shock 
of SNR RX~J1713.7$-$3946, which is likely to interact with 
cloud~A, as evidenced by   
a very high ratio of CO($J$=2--1)/CO($J$=1--0) \citep{Butt01}.
The particles can be accelerated in the shell  of   RX~J1713.7$-$3946 and 
afterwards enter cloud~A, given that the shell 
is a certain site of particle acceleration, which follows from
synchrotron X-ray emissions.  It is possible 
that particle acceleration also takes place at the secondary shocks inside
(or in the vicinity of) cloud~A initiated by the main shock of the SNR.
Finally, if there is no \emph{physical} link between cloud~A and  
SNR RX~J1713.7$-$3946,  this would imply the existence of 
unseen internal accelerators inside the molecular cloud. 
Although we cannot exclude any of these possibilities, 
the preferred option seems to be the 
``cloud interacting with SNR'' scenario.  

Formally, there are also several options for the nonthermal 
X-ray production mechanisms: synchrotron radiation, inverse 
Compton (IC) scattering, and electron and/or proton bremsstrahlung.  
However, here more definite conclusions can be drawn.

\subsection{Difficulties in Synchrotron and IC Processes}

Synchrotron X-ray emission by multi-TeV electrons
could not explain the flat  X-ray spectrum.
Because of the fast synchrotron cooling on a timescale of
$\tau_{\rm syn} \sim 150\ (\varepsilon /5~{\rm keV})^{-1/2}$ yr
(for a magnetic field of $25\,\mu$G typical in molecular clouds), 
the photon index of the synchrotron spectrum 
at X-ray energies becomes $\Gamma=\alpha_{\rm e}/2+1$, where 
$\alpha_{\rm e} \sim$ 2--2.2 is the electron acceleration index.
Thus, the differential spectrum of synchrotron X-radiation
cannot be flatter than $\varepsilon^{-2}$. 
Note that the power-law fits of the X-ray spectra of regions 2--4
($\Gamma =$ 2.1--2.3) agree with the 
synchrotron origin of X-rays from these regions.

Inverse Compton scattering is
precluded by low brightness of synchrotron radio emission.
Relatively low-energy (GeV) electrons that produce X-rays  
via  scattering off the cosmic microwave background photons, 
also emit synchrotron radiation at radio frequencies. Thus,  
assuming that the X-ray flux is due to IC scattering,
we can calculate the radio flux density of the associated synchrotron 
emission.  For a magnetic field of $25\,\mu$G the latter is expected 
at a level of $10^4$ Jy at  843 MHz, which is higher, 
by three orders of magnitudes, than the upper limit on the radio flux 
from the direction of cloud A  \citep{Slane99}.
Therefore, we can safely exclude the IC origin of the observed X-radiation.

\subsection{1/$\varepsilon$ Bremsstrahlung} \label{sec:brems}

X-rays can be produced through the bremsstrahlung of 
nonthermal electrons and/or protons interacting with the cloud gas.  
The energy distribution, for either accelerated  
protons or electrons within the cloud, is characterized 
by the position of the break energy,  $E_{\rm br}$, which is set by equating 
the lifetime against the ionization losses and
the age of the accelerator,  $\tau_{\rm ion} = \tau_{\rm age}$.
At energies below $E_{\rm br}$, for which $\tau_{\rm ion} < \tau_{\rm age}$,
the particle distribution becomes ``loss-flattened'' due to 
ionization losses \citep{Uchiyama02}. As a result, the bremsstrahlung 
spectrum obey a characteristic energy distribution
close to  $I(\varepsilon) \propto 1/\varepsilon$, 
provided that the energy spectrum of the subrelativistic particles 
becomes harder than $E^{-1/2}$.
Almost independent of the details of the acceleration spectrum,
this criterion could  be satisfied by the loss-flattened distribution
below $E_{\rm br}$,
which is controlled by the density of the cloud $n$
and the  source age $\tau_{\rm age}$.
For $n =10^3\ \rm cm^{-3}$ and $\tau_{\rm age} = 10^3$ yr,
the break appears at $E_{\rm p,br} \simeq 20$ MeV in the proton distribution, 
and correspondingly at 
$\varepsilon_{\rm p,br} = (m_{\rm e}/m_{\rm p}) E_{\rm p,br} \simeq 11$ keV 
in the proton bremsstrahlung spectrum.
On the other hand, the breaks are 
$E_{\rm e,br} = \varepsilon_{\rm e,br} \simeq 7$ MeV for electrons.
At photon energies anywhere below $\varepsilon_{\rm br}$, 
the characteristic $1/\varepsilon$ power-law is predicted.
Note that the $1/\varepsilon$ bremsstrahlung 
photons below $\varepsilon_{\rm br}$ are produced predominantly 
by particles in  a narrow energy interval  around $E_{\rm br}$.
The spectral shape of the flat X-ray spectrum of AX~J1714.1$-$3912
agrees perfectly with the $1/\varepsilon$ emission of 
either proton bremsstrahlung (PB) or electron bremsstrahlung (EB)
from the loss-flattened distribution.

Whereas the PB and EB mechanisms are expected to
give rise to the same $1/\varepsilon$ X-ray spectrum,
the PB luminosity peaks at (hard) X-rays and 
the EB peaks at gamma-rays for typical parameters.
In the case of the $1/\varepsilon$ EB emission,
the photon flux per each logarithmic interval in photon energy
anywhere up to $\varepsilon_{\rm e,br} \sim 7$ MeV is constant.
Then, the observed X-ray flux of $4\times 10^{-3}$ photon cm$^{-2}$ s$^{-1}$ 
in the 1--10 keV band implies that the 1--10 MeV flux should be
comparable to the Crab. This is inconsistent with 
non-detection by the COMPTEL instrument \citep{Schoenfelder96}
onboard the Compton Gamma Ray Observatory.
Moreover, unless we introduce a sharp cutoff in the electron distribution  
at an energy of around 100 MeV, the EB model
overshoots the $\gamma$-ray flux reported by the EGRET instrument. 
Therefore, we may conclude that  the gamma-ray data 
favor proton bremsstrahlung, 
rather than the electron bremsstrahlung scenario.  

\subsection{Energetics Requirement of 1/$\varepsilon$ Bremsstrahlung} 

The X-ray luminosity of the $1/\varepsilon$ bremsstrahlung emission
of either protons or electrons can be estimated as 
\begin{equation} \label{eq:L_x}
L_X \sim \eta_{\rm p,e} \left( \frac{W_{\rm p,e}}{\tau_{\rm age}} \right) ,
\end{equation}
where $W$ is the total amount of kinetic energy 
contained in X-ray emitting particles, $\tau_{\rm age}$ is the age 
of the source (accelerator), and $\eta=\tau_{\rm ion}/\tau_{\rm brem}$.
Here, we take into account that the bulk of X-rays are produced 
by particles from the break region, thus $\tau_{\rm age}=\tau_{\rm ion}$. 
For the parameters chosen above, $\eta_{\rm p}=3 \times 10^{-5}$
and $\eta_{\rm e}=9 \times 10^{-5}$.
These factors demonstrate the low efficiency of the 
bremsstrahlung mechanism  relative to the ionization losses,
implying a huge,  $\dot{W}_{\rm p}=5.6 \times 10^{39} d_6^2$ erg~s$^{-1}$
and $\dot{W}_{\rm e}=1.9 \times 10^{39} d_6^2$ erg~s$^{-1}$,
injection power in protons and electrons, respectively. 
Correspondingly,
the total amounts of energy  released during operation of 
the accelerator are 
$W_{\rm p}^{\rm sub} = 
1.8 \times 10^{50} d_6^2 (\tau_{\rm age}/10^3 {\rm yr})$ erg  
and $W_{\rm e} = 7.2 \times 10^{49} d_6^2 (\tau_{\rm age}/10^3 {\rm yr})$ erg. 

If the particles are accelerated in the shell of SNR 
RX~J1713.7$-$3946, and only a relatively small (10\% or so) 
fraction of these particles enter cloud A, 
the total kinetic energy, which is transferred to 
subrelativistic protons, should be at least $10^{51}$ erg;
the electron bremsstrahlung model alleviates
the total energy by a factor of about 3.
Given the limited energy budget of a supernova shock
of about $10^{51}$ erg, we face a serious problem 
to support the required X-ray luminosity,
unless the system is closer to the Earth than estimated
by the radial velocity of cloud A. For example, a distance 
of 2 kpc would reduce the energy requirement to a 
quite comfortable level of $\sim 2 \times 10^{50}$ erg. 
Several other scenarios may be invoked to overcome  
the energy budget  problem. The simplest assumption would be 
that the explosion energy significantly exceeds $10^{51}$ erg, 
and that more than 
10\% of this energy is released in the form of subrelativistic particles.
It is also possible that the nonthermal particles are accelerated 
inside the cloud, e.g.  by the shock initiated at the collision of the blast 
wave of RX~J1713.7$-$3946 with cloud A \citep{Bykov00}.
Finally, we may  speculate that
the essential part of the energy which goes into the
ionization and heating of gas, and also the excitation of plasma 
waves, is returned  to subrelativistic protons 
through acceleration on these plasma waves.
Apparently, all of these assumptions need detailed quantitative studies.

AX~J1714.1$-$3912 and cloud A are located within the error box of 
the unidentified $\gamma$-ray source 3EG J1714$-$3857 \citep{Hartman99}.
The GeV $\gamma$-ray flux may be explained by $\pi^0$ gamma-rays
by collisions between the shock-accelerated protons and cloud~A.
\citet{Butt01} argued that an electron-bremsstrahlung origin for the 
GeV flux is less likely because of the faint synchrotron radio 
emission of cloud~A.
The total energy liberated in $\gamma$-rays through the decay  
of neutral pions is roughly estimated to be
$L_{\gamma} \sim (1/3) \sigma_{\rm pp} n c W_{\rm p}^{\rm rel}$
where $\sigma_{\rm pp} \simeq 30$ mb is 
the proton--proton inelastic cross-section,
and $W_{\rm p}^{\rm rel}$ is the energy content in GeV 
protons, within cloud~A.
The EGRET luminosity,
$L_{\gamma} = 6.5\times 10^{35} d_6^2$ erg~s$^{-1}$,
corresponds to 
$W_{\rm p}^{\rm rel} = 
2.2 \times 10^{48} d_6^2 (n/10^3 {\rm cm}^{-3})^{-1}$ erg.
Thus, with the proton-bremsstrahlung model for AX~J1714.1$-$3912,
the energy content in subrelativistic protons 
far exceeds that in relativistic protons, 
$W_{\rm p}^{\rm sub} \simeq 80 \  W_{\rm p}^{\rm rel}$.
In this case, it would appear that the acceleration/injection mechanisms 
allow a small fraction of protons to be accelerated to relativistic energies.
Another possibility   could be that  the bulk of accelerated 
relativistic protons have already diffused away,
while the subrelativistic protons are still being captured within the 
molecular cloud. 

%%%%%%%%%%%%%%%%%%%%%%%%%%%%%%%%%%%%%%%

\vspace{2mm}

We thank V. Dogiel for useful discussions.
Y.U. is supported by the Research Fellowships of
the Japan Society for the Promotion of Science for Young Scientists.

%%%
% See the manual for the detail.
%%%

\end{document}